\documentclass{sva}    % onecolumn (standard format)
\smartqed  % flush right qed marks, e.g. at end of proof
\usepackage{graphicx}
\usepackage{setspace}
\usepackage[utf8]{inputenc}
\usepackage{color}
\usepackage{amsmath,amssymb}
\usepackage[colorlinks=true, urlcolor=blue, pdffitwindow=true, linkcolor=blue, citecolor=blue, pdfstartview={FitH}]{hyperref}
\usepackage{epstopdf}
\usepackage[square, sort&compress]{natbib}
\usepackage{balance}
\begin{document}
\title{\Large Existence of quantum states for Klein-Gordon particles based on exact and approximate scenarios with pseudo-dot spherical confinement}
\titlerunning{Exact and approximate solutions of KG equation}       
\author{Sami Ortakaya    %etc.
}
\institute{Shipito Address \at
             444 Alaska Avenue Suite $\#$BKF475, Torrance 90503, CA, USA \\
              %Tel.: +123-45-678910\\
              %Fax: +123-45-678910\\
              \email{sami.ortakaya@yahoo.com}             \\
             \emph{Present address: Ercis Central Post Office, 65400, Van, Turkey}  %  if needed
          % \and
          % S. Author \at
          %    second address
}
\date{\color{blue}\today}
%\vspace*{-50pt}
\maketitle
\begin{abstract}
In the present study, Kummer's eigenvalue spectra from a charged spinless particle located at spherical pseudo-dot of the form $r^2+1/r^2$ is reported. Here, it is shown how confluent hypergeometric functions have principal quantum numbers for considered spatial confinement. To study systematically both constant rest-mass, $m_{0}c^2$ and spatial-varying mass of the radial distribution $m_{0}c^2+S(r)$, the Klein-Gordon equation is solved under exact case and approximate scenario for a constant mass and variable usage, respectively. The findings related to the relativistic eigenvalues of the Klein-Gordon particle moving spherical space show the dependence of mass distribution, so it has been obtained that the energy spectra has bigger eigenvalues than $m_{0}=1$ fm$^{-1}$ in exact scenario. Following analysis shows eigenvalues satisfy the range of $E<m_{0}$ through approximate scenario.

% \PACS{PACS code1 \and PACS code2 \and more}
% \subclass{MSC code1 \and MSC code2 \and more}
\end{abstract}
\large
\section{Introduction}
Quantum mechanical wave functions are represented by probability distributions near a certain spatial point which are localized in the interaction field. Based on the spatial motions for quantum mechanical particles--represented by relativistic and nonrelativistic eigenstates-- it is important to analyze the discrete energies for quantum systems in the electronic, nuclear and particle physics. In addition to numerous studies, the quantum physical process has been applied to the external field on the electronic-interactions through plasma \cite{Falaye2019} and condensed matter \cite{Ghazi2015, shankar, impurity, dirac}. Concerning the Klein-Gordon equation, which describes relativistic spin-zero energy levels, it has been shown that the eigenvalue equation leads to spatial confluent hypergeometric functions, not only in the harmonic oscillator \cite{ikh, Sami2012, mirac} , but also in the fractional regime \cite{Tapas2020}. Furthermore, Mie-formation \cite{mirtaleb}, exponential variables \cite{ikot} and the non-central oscillatory \cite{ahmadov} have been solved for the equality on the radial distributions of the rest-mass energy. Within framework of the Klein-Gordon oscillator, commutative \& non-commutative cases and \cite{cuzinatto}, scenario of Lorentz-violating \cite{fahmed2} have been also analysed. Regarding 1D-quantum well, tunnelling \cite{tunel1} and deep well \cite{tunel2, tunel3, tunel4} has been studied for spin-0 regime. 

Besides typically eigenvalue equations  for nonrelativistic context, the spin-zero relativistic minimal form is given in the following Klein-Gordon equation \cite{Greiner}
\begin{equation}\label{h2}
\big[-\nabla^2+M^2\big]\psi_{n}(\vec{r})=\big[E_{n}-V(r)\big]^2\psi_{n}(\vec{r})
\end{equation}
where, $E_{n}$ is energy eigenvalue, $V(r)$ denotes spatial dependent potential energy, $M$ is rest mass energy of the particle system in atomic units ($\hbar=c=1$). The potential energy for the quantum mechanical particles subject to the interaction forces, plays a key role on the variable differential equations. In particular, for the solution of quantum mechanical wave equations in the defined space, the Frobenius method for spin-0 scalar particles \cite{fahmed}, the asymptotic iteration method within the scope of the molecular oscillator \cite{brzo}, and the Nikiforov-Uvarov method on the thermodynamic concepts \cite{thermo} have been used. These methods have been pioneer concepts in expressing the explicit form of the energy spectrum and the corresponding polynomial wave functions. Additionally, supersymmetric quantum mechanics has been also employed in relativistic calculations \cite{onate}.

Another analytical approach via definition of the spatial-domain is the Laplace integral transformation, which demonstrates the dependence of the energy spectra on the quantum numbers. This approach has been used not only in time-dependent problem \cite{riahi}, but also spatial part of Schrödinger equation \cite{refchen}, which is considered within exponential variables when obtaining in the $s$-domain of the Laplace transform. Here, the binomial form of the transformed space has been introduced using a multi-valued context. The studies which follow the Laplace transformation, involve the applications of spin-0 particles \cite{comm} and Dirac's spinor-systems \cite{aop} with Morse oscillator. Additionally, the \textit{N}-sphere system has been examined via reduced form in the context of spin-0 particles through pseudoharmonic oscillator \cite{nchin}. These approaches mainly assume spatial varying mass; however, comparative \& reasonable analysis including the approximate results near spatial-point is needed on the effective potential energies, so I introduce to solve spin-0 regime with familiar "equality" between scalar and vector potential (see Ref. \cite{fleischer}). I also focus on the representation of the quantum states for spin-0 particles with constant rest mass in the effective potential energies.

The purpose of this study is to model the relationship between the hypergeometric functions and the Laplace transform method, as seen in previous studies \cite{comm, aop, nchin, conv}, and is to review relativistic spin-0 eigenvalue spectra. Specifically, I aim to demonstrate how the key properties of the real function in the Laplace's $s$-domain lead to principal quantum numbers, so Kummer's differential equation through algebraic equation is revisited. In order to show reducing to solvable regime, two considerations of the Klein-Gordon equation with pseudo-dot confinement is followed: The eigenvalues provide an \textbf{exact scenario} when mass-distribution is $M=m_{0}+S(r)$ under $V(r)=S(r)$ and I will show  that the eigenvalues lead to \textbf{approximate cases} due to $r^4$ and $r^{-4}$ when constant mass is $M=m_{0}$ under $S(r)=0$. The approximate solutions for a constant rest mass $M=m_{0}$ can be illustrated on the Klein-Gordon equation, which can be reduced to Schrödinger-type equations. For this purpose, I show the "existence of quantum numbers" in the Kummer's differential equation. Within the transformed Klein-Gordon equation, I deal with a constant mass phenomena which cause to high-order power of the spatial variables \cite{samiarxiv}. As we know the spatial variables in the Laplace's $s$-domain, the Kummer-type equations define easily the eigenvalue-spectra through central potential \cite{refchen}. The radial part of the original function can be transformed into Laplace's $s$-domain which provides terminal value theorem  in a real statement \cite{schiff}.
In this way, a revised model within the analytical framework of the spinless relativistic energy spectra given in Eq. (\ref{h2}) is presented. As will be seen, the physical wave function and existence of eigenvalue spectra are given in following Section. 
In the rest of the paper, the relativistic spin-zero scheme is found by introducing that numerical results.
\section{Mathematical Statement on the Existence of Quantum Numbers}\label{Sec:results}
Considering the eigenvalues of levels $E_{n}$ and corresponding distribution of $|\psi_{n}(\vec{r})|^2$, the radial differential equation of this eigenvalue equation is given as 
\begin{equation}\label{maineq}
a(x)u''(x)+b_{n}u'(x)+c_{n}(x)u(x)=0,
\end{equation}
where $u(x)$ is unknown real function related to $E_{n}$ eigenvalue, $a(x)$, $b_{n}$ and $c_{n}(x)$ are first variable, constant term and spatial function, respectively. The eigenvalues also appear in both $b_{n}$ and $c_{n}(x)$; moreover, I take the spherical space of the two separated part $\psi_{n}(\vec{r})=u(r)Y(\theta,\,\varphi).$ 
The solution of Eq. (\ref{maineq}) may show a distribution in a certain "small" range within the Kummer-type differential equation  which also showing multivalued functions.
In a way, we need to analyze the existence of corresponding quantum numbers which is called by eigenvalue spectra. 
\\{}\\
\noindent{\bf Definition 1.} Considering a particle of wave function $\psi(\vec{r})$, Schrödinger type $n$-eigenvalue equation is given in following $N$-spherical equation
\begin{equation}\label{ndek}
\psi_{rr}+\frac{N-1}{r}\psi_{r}-\frac{\hat{L}}{r^2}\psi+\lambda_{n}(E_{n},\,r)\psi=0,\qquad\psi(\vec{r})\in (0,\,\infty)
\end{equation}
where $\hat{L}$ denotes hyperangular-momentum operator provides the hyperspherical harmonics of the function $Y(\theta_{1},\,\theta_{2},\,\theta_{3},\,\dots\theta_{N-2},\,\varphi)$ in $N$-spheres. $\lambda_{n}(E_{n},\,r)$ represents a central function combined with eigenvalue and spatial dependent potential through $\Arrowvert\psi(\vec{r})\Arrowvert=1$ and $N\geq 3$. 
\\{}\\
\noindent{\bf Definition 2.} In the space of the range $r\in (0, \infty)$, the separated radial wave function of $\psi_{n}(\vec{r})$ is a distribution $u(x)$ which satisfies dimensionless regime of variable $r\to x$. The eigenvalues provide that
\begin{equation}\label{ux}
xu''+\beta_{0}u'+\left(\beta_{1}-\beta_{2}^2 x-\frac{\beta_3^2}{x}\right)u=0, \qquad u(x)\in (0, \infty).\;
\end{equation}
Here, $\beta_{i}$ denote constants including eigenvalues and other parameters via the condition of $\Arrowvert u(r)\Arrowvert=1$ for $\psi(\vec{r})=u(r)Y(\theta,\,\varphi)$ in $N=3$. 
\subsection{Rearranging of Parameters and Variables}
The key feature is to get the following ansatz solution of transform \begin{equation}\label{ansatz}
u(x)=x^{-|\sigma|}f(x),\qquad \sigma \in \Re
\end{equation} so Equation (\ref{ux}) reduces to a kind of Kummer's equation which is given by
\begin{equation}\label{maine0}
xf''(x)+\beta f'(x)+\left(\beta_{1}-\beta_{2}^2 x\right)f(x)=0,\qquad \beta=\beta_{0}-2|\sigma|.
\end{equation}
Then we obtain that
\begin{equation}
    |\sigma|=-\frac{1-\beta_0}{2}+\sqrt{\left(\frac{1-\beta_0}{2}\right)^2+\beta_3^2}.
\end{equation}
As will be seen in the variable covering of the transformation $r\to x$, we should have $\beta_{0}=1$ through $\beta=1-2\beta_3$ for $|\sigma|=\beta_{3}$. Here, note that the given values of $\beta,\,\beta_{0}$ and $\sigma$ are valid for variable $x\propto r$. Here, we can combine these values provide special cases under $x\propto r^{\alpha},\; (\alpha=1,\,2,\,3,\,\dots)$. As can be seen in the previous solutions \cite{nchin}, it has been obtained that $\beta_0 =\frac{N}{2}$ for $\alpha=2;\,x\propto r^2$. 

We have to also consider that $f(x)$ is a well-behaved function through physical acceptable solutions: \begin{equation}u(x)=x^{\sigma_0-|\sigma|}g(x),\qquad(\sigma_{0}-|\sigma|>0)\end{equation}
which is provided by the radial boundary-values
$u(0)=0$ and $u(x\to\infty)\to 0$. After the acceptable results depend on eigenvalue equations which include components asymptotically, Equation (\ref{ux}) can be transformed into the Kummer's differential equation in the dimensionless form   
\begin{equation}\label{maineq2}
xh''+(b_{n}-x)h'-a_{n}h=0,\qquad u(x)=x^{\sigma_{0}-|\sigma|}{\rm e}^{-\beta_{2} x}h(x).
\end{equation}
Here, $b_n$ and $a_{n}$ include eigenvalue of the operator for given values of $\beta_{i}$ in Eq. (\ref{ux}). One of the solutions of Eq. (\ref{maineq2}) is confluent hypergeometric function including rising factorial. Then, the polynomial solution is
\begin{equation}\label{maineq3}
h(x)=M(a_{n},\,b_{n},x)=\sum_{{j=0}}^{\infty}\frac{a_{n}^{(j)}x^j}{b_{n}^{(j)}j!}.
\end{equation}
As I will proof, the eigenvalue spectra leads to
\begin{equation}\label{integer}
a_{n}=-n,\qquad n=0,\,1,\,2,\,3,\,\dots
\end{equation}
The proposed wave functions have to be physical boundaries, so spherical regime is provided to be effective way in keeping up with ansatz solution.\\{}\\
\noindent \textbf{Lemma 1.}\label{lem:11}
Let $u(x), f(x), g(x), h(x)\in  \Re$ and let $\sigma$, $\sigma_0$ be real in eigenvalue parameters. Then the following assertions hold:
\begin{enumerate}
\item If $f(x)$ is a well-behaved function providing that $u(x)$ yields a non-zero distribution. Furthermore, boundary values with spherical regime permit the our comment on the radial distributions. Then, the conditions $u(0)=0$ and $u(x\to\infty)\to 0$ denote physical acceptable solutions satisfy that 
\begin{equation}
u(x)=x^{-|\sigma|}f(x),\qquad f(x)=x^{\sigma_{0}}g(x)\;\;{for}\;\;\sigma_{0}> |\sigma|.
\end{equation}
 
\item Due to the asymptotic behaviour of the $u(x)$,  the behaviours of $f(x)$ and $g(x)$ are also built up at long distances:
\begin{equation}\label{limitler}
\lim_{x\to\infty}f(x)\to0\;\;{and}\;\; \lim_{x\to\infty}g(x)\to0.
\end{equation}
\end{enumerate}
\subsection{Solutions of Kummer's Eigenvalue Spectra}
In the presence of the spherical wave functions in Klein-Gordon Equation (\ref{h2}), we can conclude that the radial form of Equation (\ref{ux}) yields an eigenvalue dependence of variables. In addition to the closed-form ansatz in Equation (\ref{ansatz}), the terminal-value theorem and the existence of eigenvalue numbering will provide to obtain physical solution. There are two cases defined by the physical wave function with the $n^{\rm th}$ eigenvalue.\\{}\\
\noindent\textbf{Case 1.}
A kind of the Kummer's eigenvalue equation is obtained as a form of Equation (\ref{maine0}). As a different way, multi-valued functions can be analyzed by considering real functions in $s$-domain. We should have a first solution of the ordinary equation in the following from:
\begin{equation}\label{uu}
xf''(x)+\beta f'(x)+\left(\beta_{1}-\beta_{2}^2 x\right)f(x)=0, \qquad u(x)=x^{-|\sigma|}f(x),\qquad \sigma \in \Re,
\end{equation}
then Laplace's $s$-function reads \cite{refchen}
\begin{eqnarray}
\begin{aligned}
\mathcal{L}\{f(x)\}&=F(s)\\
&=A_{n}\left(s-\beta_{2}\right)^{a}\left(s+\beta_{2}\right)^{b}
\end{aligned}
\end{eqnarray}
with
\begin{equation}
    a=-\frac{2-\beta}{2}+\frac{\beta_1}{2\beta_2},\qquad b=-\frac{2-\beta}{2}-\frac{\beta_1}{2\beta_2}
\end{equation}
where $A_{n}$ is a determined constant by putting inverse transform, which includes new constant $C_n$ in the eigenfunctions which consist of confluent hypergeometric functions:
\begin{equation}\label{integer2}
f(x)=C_{n}{\rm e}^{-\beta_{2}x}x^{1-\beta_{0}+2|\sigma|}M(-a,\,2-\beta,\,2\beta_{2}x),\qquad a=n,\quad (n=0,\,1,\,2,\,3,\dots).
\end{equation}\\{}\\
\noindent{\it Proof.} The Eq. (\ref{uu}) yields an ordinary differential equation in Laplace's $s$-variable, which is obtained in following form:
\begin{equation}
    (s-\beta_{2}^2)F'(s)+\left[(2-\beta)s-\beta_{1}^2\right]F(s)=0.
\end{equation}
 On the other hand, $s$-domain functions provide that $F(s)\in\Re$ at $s=0$.
Then we have real function of the form $$F(0)=(-1)^{a}\beta_{2}^{\frac{a+b}{2}}$$
and then we should get the eigenvalue spectra \begin{equation}\label{integer3}
a=n,\qquad n=0,\,1,\,2,\,3,\,\dots
\end{equation}
Note that the terminal-value theorem is valid for real values of spatial wave function $f(x)$.
Inverse transform also yields convolution integral through solutions \cite{schiff}
\begin{eqnarray}
\begin{aligned}
\mathcal{L}^{-1}\{F(s)\}&=f(x)\\
&=B_{n}\int_{0}^{x} (x-\tau)^{-a-1}\tau^{-b-1} {\rm e}^{2\beta_2 \tau}{\rm d}\tau
\\&=C_{n}{\rm e}^{-\beta_2 x}x^{1-\beta}M\left(-a,\,2-\beta,\,2\beta_2 x\right),\quad \beta=\beta_{0}-2|\sigma|
\end{aligned}
\end{eqnarray}
where $B_{n}$ denotes eigenvalue dependent constant including Gamma function \cite{stegun}. One can see that the acceptable wave function is provided by $f(0)=0$ and convergence limit which is given in Equation (\ref{limitler}). Note that the well-behaved distribution of the function $f(x)$ converges with $sF(s)$ in Equation (\ref{terminal}).\qed
\begin{center}\end{center}
\noindent\textbf{Case 2.}
Equation (\ref{ux}) denotes another kind of the Kummer's eigenvalue equation
\begin{equation}
xh''(x)+(\varepsilon_{1}-\varepsilon_{2} x)h'(x)+\varepsilon_{3}h(x)=0,
\end{equation}
where
\begin{equation} \varepsilon_1=2|\sigma|+\beta_0,\quad\varepsilon_2=2\beta_2,\quad\varepsilon_3=\beta_{1} - (2|\sigma|+\beta_0 ),
\end{equation}
then $h(x)$ provides that the confluent hypergeometric functions related to the $n$'th eigenvalues:
\begin{equation}
h(x)=C_{n} M(-n,\,\varepsilon_1,\,\varepsilon_{2}x),\qquad \frac{\varepsilon_3}{\varepsilon_2}=n,\quad n=0,\,1,\,2,\,3,\,\dots
\end{equation}\\{}\\
\noindent{\it Proof.} Applying the Laplace's transform in $s$-domain, one can obtain an eigenfunction of the following form:
\begin{eqnarray}
\begin{aligned}
\mathcal{L}\{h(x)\}&=F(s)\nonumber\\&=A_{n}\left(s-\varepsilon_{2}\right)^{\varepsilon_{1}-2}\left(\frac{s-\varepsilon_{2}}{s}\right)^{1+\frac{\epsilon_{3}}{\epsilon_{2}}},
\end{aligned}
\end{eqnarray}
where $A_{n}$ is a determined constant. In the case of the terminal-value, we should consider that
\begin{equation}\label{terminal2}\lim_{x\to\infty}h(x)=\lim_{s \to 0}sF(s)\to \infty\qquad s\in\Re,\end{equation}
which satisfies that
\begin{equation}\label{terminal3}\lim_{x\to\infty}u(x)\to 0,\end{equation}
and then we conclude that the "zero and positive integers" at $s=0$. This condition is provided by
$$\frac{\varepsilon_3}{\varepsilon_2}=n,\quad n=0,\,1,\,2,\,3,\,\dots$$
Here, one can see that real function is obtained via the integers related to $s<\varepsilon_{2}$. We also obtain that the convolution integral yields \cite{aop}
\begin{eqnarray}
\begin{aligned}
\mathcal{L}^{-1}\{F(s)\}&=h(x)\nonumber\\&=C_{n} M(-n,\,\varepsilon_1,\,\varepsilon_{2}x).
\end{aligned}
\end{eqnarray}\qed

\noindent Here, we expect that the radial function of transformation $f(x)$ shows an agreement via terminal-value theorem in Laplace $s$-domain. Then, real values represent integer-context with exponent in combined with following theorem:\\{}\\
\noindent{\bf Theorem.} (\cite{schiff}) Suppose that $f(x)$ satisfies
the conditions of the derivative theorem and furthermore that
$\lim_{x\to \infty} f (x)$ exists. Then this limiting value is given by \begin{equation}\label{terminal}\lim_{x\to\infty}f(x)=\lim_{s \to 0}sF(s)\qquad s\in\Re,\end{equation}
where $F(s)=\mathcal{L}\{f(x)\}$.

\section{Numerical Results}\label{sec1}
The obtained values are resulting in the pseudo-dot structure when mass parameters are to be $m_{0}c^2$ and $m_{0}c^2+V(r)$ for exact and approximate scenario, respectively. 
There is no magnetic field and variable of the first component according to pseudo-dot energy can be given via following function
\begin{equation}
    V(r)=D_{\rm e}\left(\frac{r}{r_0}-\frac{r_{0}}{r}\right)^2 ,
\end{equation}
where $D_{\rm e}$ and $r_0$ are energy value of quantum well-width and turning point related to separation, respectively.
The exact solutions are valid for the Klein-Gordon equation in Eq. (\ref{h2}) which reduced to Schrödinger-type equation as following form
\begin{equation}\label{h3}
\bigg[\nabla^2+E^2-m_{0}^2-2(E+m_{0})V(r)\bigg]\psi_{n}(\vec{r})=0
\end{equation}
Here, mass distribution is taken by $M=m_{0}+V(r)$. Within framework of the approximate scenario, it has been also obtained that the Klein-Gordon equation reads
\begin{equation}
\label{h4}
\bigg[\nabla^2+E^2-m_{0}^2-2EV(r)+
V^2 (r)\bigg]\psi_{n}(\vec{r})=0 ,  
\end{equation}
here rest-mass energy is taken constant in the form, $M=m_{0}$. Putting pesudo-dot into Eq. (\ref{h4}), forth orders are obtained via $r^4 + 1/r^{4}$, and then I propose an approximation eith Taylor expansion near $r_{0}$. Note that Eqs. (\ref{h3}) and (\ref{h4}) also lead to Schrödinger's form
\begin{equation}
\label{ef1212}
\bigg\{\nabla^2+\big(\epsilon-\Phi\big)\bigg\}\Psi(\vec{r})=0,
\end{equation}
where energy levels occur in view of the variable mass and constant one:
 \begin{equation}
    \epsilon_{n}=
    \begin{cases}
      E_{n}^2-m_{0}^2, & \text{variable mass} \\
      E_{n}^2-m_{0}^2+4E_{n}D_{\rm e}+6D_{\rm e}^2, & \text{constant case}
    \end{cases}
  \end{equation}
Also, effective potential can be obtained as following form \begin{equation}
   \Phi=
    \begin{cases}
     (E_{n}+m_{0})V(r), & \text{variable mass} \\
      D_{\rm e}(2E_{n}+4D_{\rm e})\left(\frac{r^2}{r_{0}^2}+\frac{r_{0}^2}{r^2}\right)-D_{\rm e}^2\left(\frac{r^4}{r_{0}^4}+\frac{r_{0}^4}{r^4}\right), & \text{constant case}
    \end{cases}
  \end{equation}  
  \begin{figure}[!hbt]
\centering
{\scalebox{.8}{\includegraphics{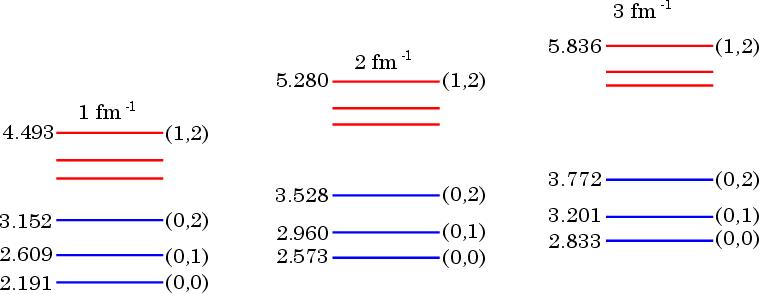}}
\scalebox{.3}{\includegraphics{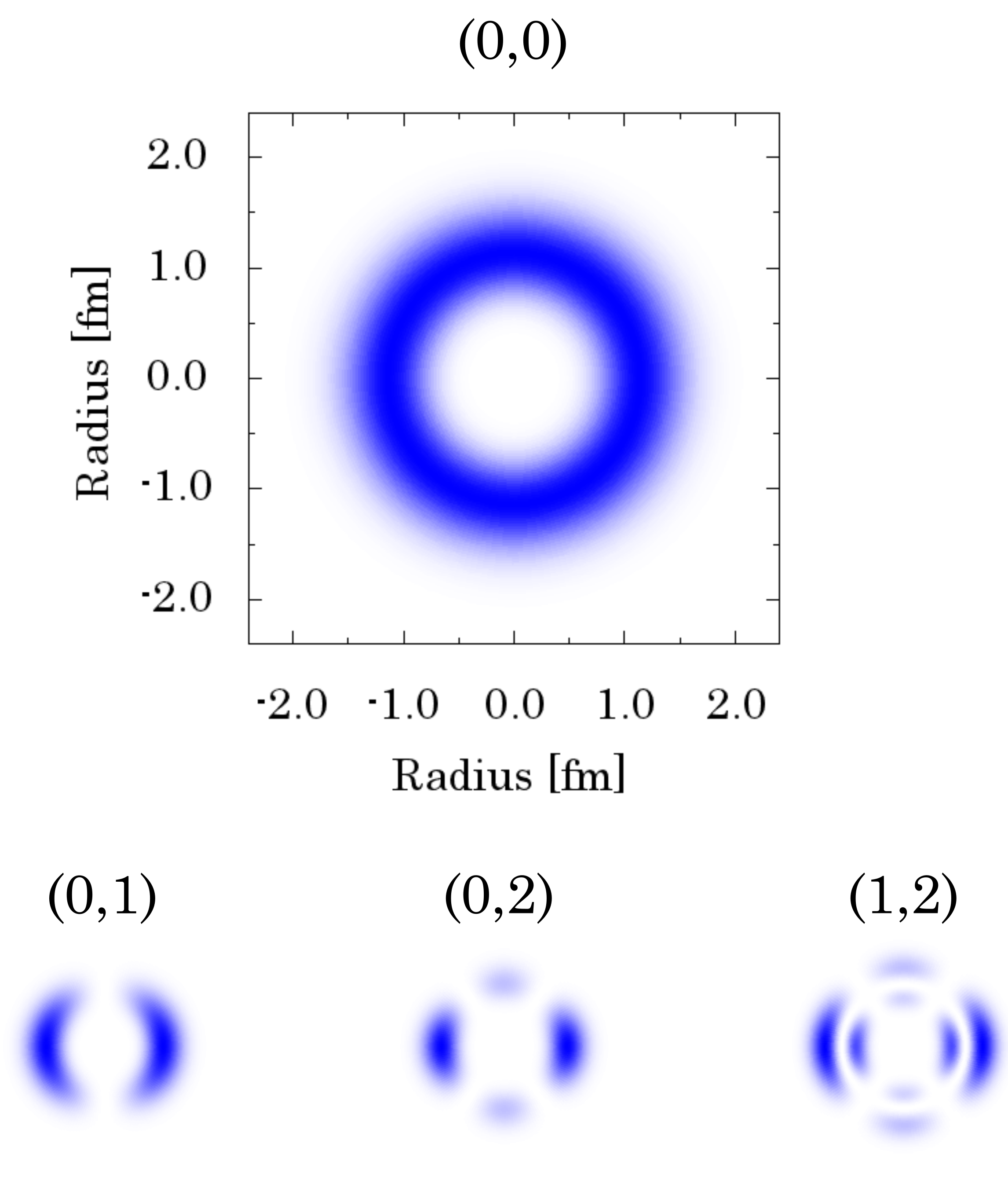}}}
\caption{Relativistic $(n,\ell)$ energy-level values and densities via exact case (in $\rm fm^{-1}$) when radial mass-distribution energy equals to scalar potential.}\label{fig1}
\end{figure}

\begin{figure}[!hbt]
\centering
\scalebox{.85}{\includegraphics{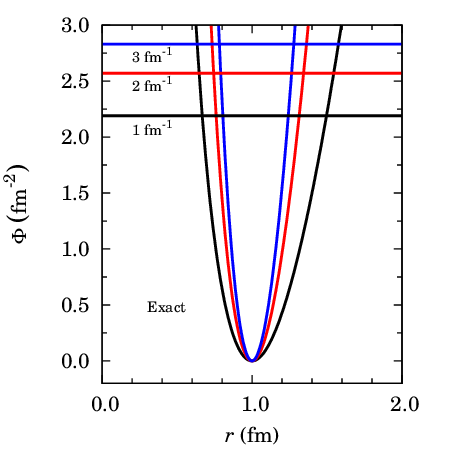}}
\caption{Effective potential energy and corresponding eigenvalues with various width parameters of $D_{e}=1,\,2\text{ and }3\, \text{fm}^{-1}$.}\label{fig2}
\end{figure}

From Eqs. (\ref{ndek}) and (\ref{ux}), we should have a fonction as $\beta_3 (E_{n},\,\ell)$ for $\ell=0,\,1,\,2,\,3,\,\dots$ . Due to the centrifugal term $\frac{\ell(\ell+1)}{r^2}$, obtained new eigenvalues are denoted by $(n,\,\ell)$. Assuming rest mass energy ($\hbar=c=1$) and separation-distance are taken through values of $m_0 =1.0$ $\rm fm^{-1}$ and $r_{0}=1.0$ fm, Figure \ref{fig1} shows the exact energy eigenvalues of relativistic spin-zero particles at values of well width parameters $D_{\rm e}=1$, 2 and 3 fm$^{-1}$. With increasing $D_{\rm e}$ we can see that the eigenvalues rise to about 5.84 $\rm fm^{-1}$. From there, one can also know that the energy eigenvalues of excited states increase with increasing quantum numbers $n$ and $\ell$. Because of the effective Schrödinger equation given Eq. (\ref{ef1212}), energy spectra shifts to bigger values with increasing $D_{\rm e}$ which represents narrow quantum well in the effective potential energy. It can be seen that increasing energies exhibit positive values in Figure \ref{fig2}, so the Schrödinger formalism shows this behavior on the given energy-levels. Moreover, radial probability distribution via $|u_{n\ell}|^2$ including confluent hypergeometric functions at $D_{\rm e}=1.0\,\rm fm^{-1}$, can be plotted near $r_{0}=1.0 $ fm, so we have ground state is valid for maximum near 1.0 fm. The obtained results related to the densities show the spherical well exhibits the expected distributions in the presence of spinless relativistic energies. 

\begin{figure}[!hbt]
\centering
\scalebox{.75}{\includegraphics{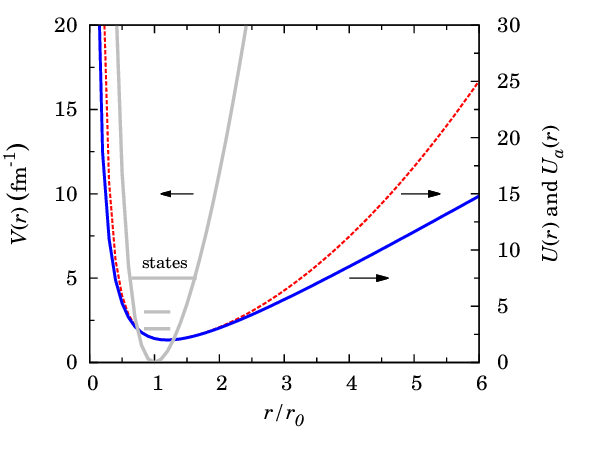}}
\caption{Behavior of the radial variable and its approximation near $r=r_{0}$.}\label{fig3}
\end{figure}

For a given constant rest mass in Equation (\ref{h2}), approximate solutions are also needed near a spatial point. Still, we can choose the same values as exact-procedure ones, so expected value lies the range $E<m_{0}$, where $\frac{m_{0}c^2}{\hbar^2 c^2}=m_{0}.$ Within the Eq. (\ref{h4}), $4^{\rm th}$ order occur when the variable of $V^2 (r)$ causes to $r^4+\frac{1}{r^4}$. Here, we can consider that 
$$U(r)=r^4+r^{-4},\qquad U_{a}(r)=A_{0}+A_{1}r^2+A_{2}r^{-2}.$$ 
These variables can be compared when $D_{\rm e}=5.0\,\rm fm^{-1}$ at $r_{0}=1.0$ fm. We can see from Figure \ref{fig3} that, besides $U(r)$ with 4th-order $U_{a}(r)=A_{1}+A_{2}x+A_{3}x^{-1}$ is also valid near $r=r_{0}$: $x=1$ with $x=r^2/r_{0}^2$ and then we can easily obtain that
\begin{equation}
x^2+\frac{1}{x^2}\simeq A_{1}+A_{2}x+\frac{A_{3}}{x},
\end{equation} 
where $A_{0}=-6$ and $A_{2}=A_{3}=4$ are obtained coefficients from Taylor expansion near $x=1$.

\begin{figure}[!hbt]
\centering
\scalebox{.8}{\includegraphics{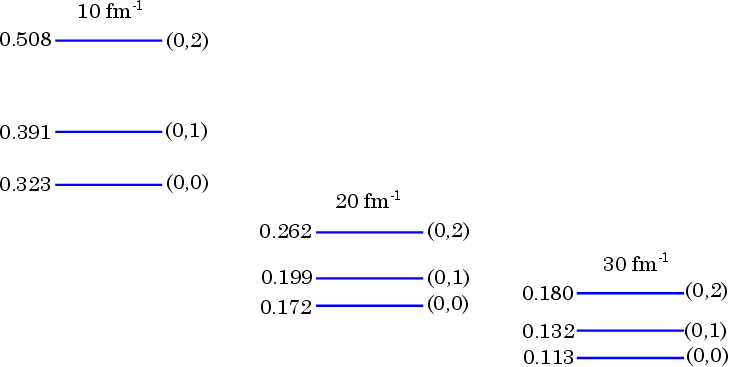}}
\caption{In the presence of constant rest mass, $(n,\,\ell)$ energy eigenvalues (in $\rm fm^{-1}$). }\label{figa}
\end{figure}

In this way, it can also be taken that the energy eigenvalues depend on approximation constants, $A_{i}$; $i=0,\,1 \text{ and } 2$. So that, similar differential equation is obtained in this argument. The eigenvalues obtained from Eq. (\ref{integer2}), yield smaller values than values of constant mass energy $m_{0}=1.0\, \rm fm^{-1}$. As can be seen in Figure \ref{figa}, these values have good agreement with $E<m_{0}$. Furthermore, numerical values under well width parameter of $D_{\rm e}=10,\,20\text{ and } 30\, \rm fm^{-1}$ decrease with increasing $D_{\rm e}$ since the considered effective potential in the approximate scenario behaves as a "quantum barrier". 

%%%%%%%%%%%%%%%%%%%%%%%%%%%%%%%%%%%%%%

\begin{figure}[!hbt]
\centering
\scalebox{.8}{\includegraphics{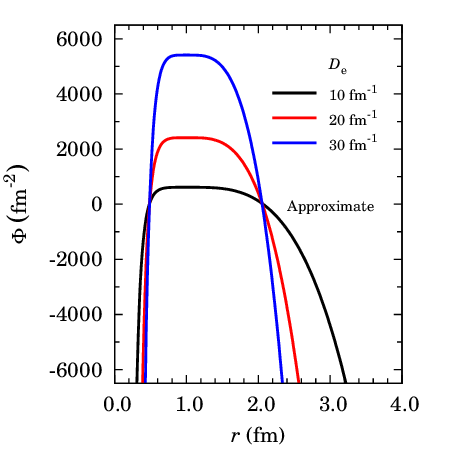}}
\caption{Effective potential energies in the approximate scenario and corresponding eigenvalues with various quantum well parameters.}\label{fig6}
\end{figure}

In Figure \ref{fig6}, we consider the wide \& narrow barriers, so we expect that the quasi-nonrelativistic eigenvalues denopted by $\epsilon_n$ increase with increasing barrier height within narrow quantum well. Here, obtained quasi values 613.894,
2414.731 and 
5414.188 $\rm fm^{-2}$ for width parameter of 10, 20 and 30 $\rm fm^{-1}$, respectively. These values corresponds to relativistic eigenvalues of 0.323, 0.172 and 0.113. 
%%%%%%%%%%%%%%%%%%%%%%%%%%%%%%%%%%%%%%%%%%%%
\section{Conclusion}\label{sec13}
In this work, the Schrödinger-type equations in the exact and approximate-solvable forms which are also represented by Klein-Gordon equation with relativistic spinless regime has been studied through Kummer's eigenvalues. As an analytical line with acceptable solutions which satisfy radial context of the range of $(0,\,\infty)$, it is analysed how Kummer's type solutions exhibit $n$-dependent solutions for $n=0,\,1,\,2,\,3,\,\dots$ . There is a key property is to get the proposed solution in a better form via
$$u(x)=x^{-|\sigma|}f(x);\qquad f(x)=x^{\sigma_{0}}g(x)\text{ for }\sigma_{0}>|\sigma|,$$
where $x$ defines radial variable in terms of the radius $r$. To best of our knowledge Mie-type variables provide that confluent hypergeometric functions even so approximation is valid near equilibrium point of the behaviour $V(r=r_{0})=0$. So that, Kummer's differential equation shows eigenvalue spectra due to the real function of Laplace transform with $s$-domain. Therefore, it can be easily obtained that the Kummer's orthogonality and eigenvalue spectra describe probability distribution in a certain spatial region.

In the presence of the Klein-Gordon equation related to the wave functions including confluent hypergeometric polynomials, two solutions which provide corresponding energy levels can be distinguished: Firstly, exact solution for pseudo-dot quantum confinement lies that $E>m_{0}$ with spinless Klein-Gordon equation. These solutions are possible through radial mass distribution $m(r)c^2=m_{0}c^2+S(r)$ via $V(r)=S(r)$. Secondly, the approximate scenario which causes to $E<m_{0}$ due to the fact that rest mass is not change (i.e, $S(r)=0$ $V(r)\neq0$) in the interval of (0, 2 fm). In a way, considered space also allows a closeness to the effective variable is obtained as $r^4+r^{-4}.$ Probability distribution is also valid at  considered range; nevertheless, it can be seen that Schrödinger-transformation under effective potential shows why eigenvalues exhibit increasing (exact form) \& decreasing (approximate scenario) with increasing well-width, so Kummer's solvable models are also used via analytical forms. 

\section{Data Availability Statement}
No Data associated in the manuscript.
\section{ Conflict of Interest}
The author declares that he has no confict of interest.

\end{document}